\documentclass[11pt]{article}
\usepackage{mathptmx}
\usepackage[pdftex]{hyperref}
\usepackage{amssymb,latexsym,fancyhdr,color,graphics,graphicx}
\newtheorem{proposition}{\bf Proposition}[section]
\newtheorem{theorem}{\bf Theorem}[section]
\newtheorem{definition}{\bf Definition}[section]

\begin{document}
\renewcommand{\theequation}{\thesection.\arabic{equation}}
\begin{center}
{\Large \bf Models of self-financing hedging strategies in illiquid markets:\\
 symmetry reductions and exact solutions}\\[2ex]
 {\large  Ljudmila A. Bordag, Anna Mikaelyan}\\[1ex]
{\it IDE, MPE Lab,\\ Halmstad University, Box 823,\\ 301 18
Halmstad, Sweden}
 \end{center}

 \noindent
{\bf Abstract}\\
We study the general model of self-financing trading strategies in
illiquid markets introduced by Sch{\"o}nbucher and Wilmott, 2000.
A hedging strategy in the framework of this model satisfies a
nonlinear partial differential equation (PDE) which contains some
function $g(\alpha)$. This function is deep connected to an
utility function.

We describe the Lie symmetry algebra of this PDE and provide a
complete set of reductions of the PDE to ordinary differential
equations (ODEs). In addition we are able to describe all types of
functions $g(\alpha)$ for which the PDE admits an extended Lie
group. Two of three special type functions lead to models
introduced before by different authors, one is new. We clarify the
connection between these three special models and the general
model for trading strategies in illiquid markets. We study with
the Lie group analysis the new special case of the PDE describing
the self-financing strategies. In both, the general model and the
new special model, we provide the optimal systems of subalgebras
and study the complete set of reductions of the PDEs to different
ODEs. In all cases we are able to provide explicit solutions to
the new special model. In one of the cases the solutions describe
power derivative products.



{\bf Corresponding author:} Ljudmila A. Bordag,
Ljudmila.Bordag@hh.se

\section{Introduction}
We study self-financing hedging strategies in the framework of  a
reaction-function model. The main subject of this model is a
smooth \emph{reaction function} $\psi$ that gives the equilibrium
stock price $S_t$ at time $t$ as function of some fundamental
value $F_t$ and the stock position of a large trader. In the
framework of this model there are two types of traders in the
market: ordinary investors and a large investor. The overall
supply of the stock is normalized to one. The normalized stock
demand of the ordinary investors at time $t$ is modeled as a
function $D(F_t,S_t)$, where $S_t$ is the proposed price of the
stock. The normalized stock demand of the large investor is
written in the form $\rho \Phi_t$; $\rho \ge 0 $ is a parameter
that measures the size of the trader's position relative to the
total supply of the stock. The equilibrium price $S_t$ is then
determined by the market clearing condition
\begin{equation}
  \label{eq:equilibrium}
  D(F_t,S_t) + \rho \Phi_t =1\,.
\end{equation}
The equation (\ref{eq:equilibrium}) admits a unique solution $S_t$
under suitable assumptions on the function $D(F_t,S_t)$. Hence
$S_t$ can be expressed as a function $\psi$ of $F_t$ and $\rho
\Phi_t$, so that $S_t = \psi(F_t,\rho \Phi_t)$.

Now we turn to the characterization of self-financing hedging
strategies in the framework of the reaction-function model.
Throughout we assume that the fundamental value process $F_t$
follows a geometric Brownian motion with volatility $\sigma$ as in
the  Black-Scholes model. Moreover, we assume that the reaction
function  $\psi(F_t,\rho \Phi_t)$ is of the form
\begin{equation}\label{reactfunct}
\psi(f,\alpha)= f ~ g(\alpha)
\end{equation}
with some increasing function $g(\alpha)$. This holds for any
model where
\begin{equation}\label{demand}
D(f,s) = U(f/s)
\end{equation}
for a strictly increasing function $U : (0,\infty)\to \mathbb{R} $
with a suitable range.

Assuming as before that the normalized trading strategy of the
large trader is of the form $\rho \Phi(t,S)$ for a smooth function
$\Phi$,  we get from It\^o's formula that
\begin{equation} \label{eq:dS-intermediate}
 dS_t = g(\rho \Phi(t, S_t))\, dF_t + \rho F_t g_\alpha(\rho \Phi(t, S_t))
 \Phi_S(t, S_t)\,d S_t + b (t,S_t)\, dt
\end{equation}
(since $S_t = g(\rho \Phi(t,S_t))F_t$). The precise form of
$b(t,S_t) $ is irrelevant to our purposes. Assume now that
\begin{equation} \label{eq:bound-phi}
 \big(1 - \rho F_t g_\alpha(\rho \Phi(t, S_t)) \Phi_S(t, S_t)\big )>0
 \quad
 \rm {a.s.}.
\end{equation}
This can be viewed as an upper bound on the permissible variations
of the large trader's strategy. Rearrangement and integration
$\big(1 - \rho F_t g_\alpha(\rho \Phi(t, S_t)) \Phi_S(t, S_t)\big
)^{-1}$ over both sides of  equation (\ref{eq:dS-intermediate})
gives us the following dynamics of $S$:
\begin{equation} \label{eq:price-reaction-function-model}
dS_t = \frac{1}{1- \rho \frac{g_\alpha (\rho \phi(t,S_t))}{g(\rho
\phi(t,S_t))} S_t \phi_S(t,S_t) } \sigma S_t dW_t +
\tilde{b}(t,S_t) dt \, ,
\end{equation}
again the precise form of $\tilde b$ is irrelevant.

If we apply the Ito-Wentzell formula \cite{bank&baum} to this
equation we obtain the following PDE for a self-financing hedging
strategy
\begin{equation}\label{MainEqL3}
u_{t}+\frac{1}{2}\frac{\sigma^{2}S^{2}u_{SS}}{\left(1-\rho\frac{g'(\rho
u_{S})}{g(\rho u_{S})}Su_{SS}\right)^{2}}=0.
\end{equation}

Under some special choices of the function $g(\alpha)$ equation
(\ref{MainEqL3}) represents the models introduced earlier.\\
If we take $g(\alpha)=c_2 e^{c_1\alpha}$, $c_1,c_2 - const.$ we
obtain the model introduced in \cite{frey}. This model was studied
with Lie group methods in \cite{bordag&frey}, \cite{bordag&chmak}
and some generalization of this model in \cite{bordag2007}.\\
In the case of $g(\alpha)=c_2(\rho+k \alpha)^{-\frac{1}{c_1}}$,
$c_1,c_2, k - const.$ equation (\ref{MainEqL3}) coincides with the
models introduced in \cite{sircar&papanicolaou} and studied with
Lie group methods in \cite{bordag2010} and \cite{bordag2009}.

The main goal of this paper is to study the analytical properties of
 equation (\ref{MainEqL3}) in the first line with the Lie group method.
We will clarify which symmetry properties has this equation and
under which additional conditions on the function $g(\alpha)$ it
admits a richer symmetry group.

If the symmetry group admitted by an equation is known then it can
be used  in many cases to reduce the given equation to a simpler
one, for instance, to an ODE or even solve it. Solutions of such
reduced equations are called invariant solutions because they are
invariant under the action of the corresponding subgroup of the
symmetry group of this equation. We follow this strategy and
describe the complete set of reductions for the general model
(\ref{MainEqL3}) in section~\ref{genSection}.

In addition we obtain also some special forms of utility functions
for which the corresponding equation admits a richer symmetry
group. The connection between the function $g(\alpha)$ and the
utility function is described in section~\ref{utility}. For a
special form of utility function  we obtain a special model for
hedging strategies in illiquid markets. This special model is
studied in section~\ref{specsl}. Also for this model we present an
optimal system of subalgebras, the complete set of reductions to
ODEs. In this case we are able to solve all the ODEs and provide
explicit formulas or graphs of solutions.

\section{The function $g(\alpha)$, the
corresponding utility function and the self-financing hedging
strategy}\label{utility}

We use the classical assumptions (see \cite{delbaen}) on a utility
function $U(x)$, modeling the utility of a market participant's
wealth $x$ at fixed time.
\begin{definition} \label{defutility}
Let $U(x)$ is a utility function. Then $U: \mathbb R \to \mathbb R
\cup \{ - \infty \}$
\begin{enumerate}
\item is increasing on
$\mathbb R $, continuous on $ \{ U > - \infty \}$, differentiable
and strictly concave on the interior of $ \{ U> - \infty \},$
\label{increasing}
\item $U'(\infty)= \lim_{x \to \infty} U'(x) =0 ,$ \label{utilityinfty}
\item if a negative wealth is not allowed then $U(x)=-\infty,$ for
$x<0$, $U(x)>-\infty,$ for $x>0$, and the Inada condition $ U'(0)
= \lim_{x \to 0, x>0} U'(x)=\infty$ holds, \label{inada}
\item if a negative wealth is allowed then $U(x)>-\infty,$ for $x \in
\mathbb R$ and $ U'(-\infty) = \lim_{x \to - \infty} U'(x)=\infty$
holds.\label{negativ}
\end{enumerate}
\end{definition}
Using the form of the factor representation of the reaction
function (\ref{reactfunct}) and (\ref{demand}) we obtain the
following equation for the utility function
\begin{equation}\label{utilityg}
U\left(\frac{1}{g(\alpha)}\right)=1-\alpha.
\end{equation}

We restrict our consideration further to the case where a negative
wealth is not allowed and consider the utility function $U$ as a
map $U: \mathbb R^+ \to \mathbb R \cup \{ - \infty \}$ with the
assumptions (\ref{increasing}) -- (\ref{inada})  listed in the
Definition \ref{defutility} above.

Because of this we can restrict the argument of the utility
function $U(x)$ on $\mathbb R^+$, hence we look for a function
$g(\alpha)>0$. We are able to rewrite the relation
(\ref{utilityg}) in a more convenient and explicit form for some
typical special forms of the function $g(\alpha)$:
\begin{enumerate}
\item If the function $g(\alpha)$ takes the form $g(\alpha)=c_2 e^{c_1\alpha}$
then it corresponds to the utility function which is equal to
\begin{equation} \label{logutility}
g(\alpha)=c_2 e^{c_1\alpha},~~ \to ~~ U(x)=\frac{1}{c_1}\ln
x+\frac{c_1+\ln c_2}{c_1},~c_1, c_2 \neq 0.
\end{equation}
Because the argument of the utility function should be positive
the function $g(\alpha)=c_2e^{c_1\alpha}$ should be positive too,
i.e. the coefficient $c_2$ should be strictly positive $c_2>0$.
According to the assumption (\ref{increasing}) of
Definition~\ref{defutility} the utility function $U(x)$ should be
an increasing function, then $c_1>0$. It is easy to prove that all
other assumptions are as well satisfied for the utility function
of the form (\ref{logutility}) by $c_1,c_2 >0$.\\
The PDE for the self-financing hedging strategy takes in this case
the following form
\begin{equation}\label{frey}
u_{t}+\frac{1}{2}\frac{\sigma^{2}S^{2}u_{SS}}{\left(1-\rho c_1 S
u_{SS}\right)^{2}}=0.
\end{equation}
This equation was first introduced in \cite{frey}. The complete
description of the symmetry group, invariant reductions and
invariant solutions to this equation as well as to some
generalization of this equation are given in \cite{bordag2007},
\cite{bordag&frey} and \cite{bordag&chmak}.

\item If the function is given by $g(\alpha)=c_2\alpha^{c_1}$
then the corresponding utility function takes the form
\begin{equation}
g(\alpha)=c_2\alpha^{c_1}, ~~ \to
~~U(x)=-(c_2x)^{-\frac{1}{c_1}}+1.
\end{equation}
Under our assumption  the utility function acts on $(0,+\infty)$
correspondingly  the function $g(\alpha)=c_2\alpha^{c_1}$ should
be positive. The assumptions (\ref{increasing}) -- (\ref{inada})
of Definition~\ref{defutility} are satisfied if
\begin{equation} \label{constdegry}
c_1, c_2 >0.
\end{equation}
The PDE which corresponds to this function $g(\alpha)$ has the
form
\begin{equation}\label{haupt}
u_{t}+\frac{1}{2}\frac{\sigma^{2}S^{2}u_{SS}u_S^2}{(u_S-c_1Su_{SS})^2}=0.
\end{equation}
This equation is the special model for hedging strategies in
illiquid markets which we will study  in the section \ref{specsl}
of this paper.

\item If we take similar to the previous case the function $g(\alpha)$ in
the form $g(\alpha)=c_2(\rho + k \alpha)^{-\frac{1}{c_1}}$, where
$c_1,c_2,k \in {\mathbb R},~~ k\ne 0$ then the corresponding
utility function is given by
\begin{equation} \label{constdegry}
g(\alpha)=c_2(\rho+k \alpha)^{-\frac{1}{c_1}}, ~~ \to ~~
U(x)=-\frac{1}{k}(c_2x)^{c_1}+\left(1+\frac{\rho}{k}\right).
\end{equation}
The utility function acts on $(0,+\infty)$ because of that the
function $g(\alpha)=c_2(\rho+k \alpha)^{-1/c_1}$ should be
positive.

The assumptions of Definition~\ref{defutility} on the the utility
function acting on $(0,+\infty)$ are satisfied if
\begin{equation}\label{potenz}
c_2>0,~~c_1 \in(-\infty,0)\cup (0,1), ~~ k c_1  > 0.
\end{equation}
The PDE for a self-financing trading strategy takes in this case
the form
\begin{equation} \label{sipa}
u_t + \frac{1}{2} \frac{\sigma^2(1+k u_S)^2 }{c_1^2 \left(1+k u_S
+ \frac{k}{c_1}
   S u_{SS}\right)^2} S^2 u_{SS}  =0 \,, {c_1} \ne 0,
\end{equation}
This equation coincides with some of the models introduced in
\cite{sircar&papanicolaou}).  This model was studied with the Lie
group method in \cite{bordag2010} and \cite{bordag2009}. The
corresponding symmetry algebra admitted by this equation, an
optimal system of subalgebras, the set of invariant reductions and
in some cases also exact invariant solutions are presented in
these papers.

\end{enumerate}

\section{The Lie algebra admitted by the general model (\ref{MainEqL3})}\label{genSection}

To study the symmetry properties of (\ref{MainEqL3}) we use the
method introduced by Sophus Lie and developed further in
\cite{ovsiannikov}, \cite{olver} and \cite{ibragimov}. For the
first reading  also the books  \cite{stephani} and
\cite{belinfante&kolman} can be used which contain many examples.

The main idea of this method can be formulated in the following
way. We introduce first a jet-bundle of the corresponding order
(in this case of the order two, because of we have to do with a
second order PDE). Then we study all smooth point transformations
which locally keep the solution subvariety of the studied PDE
invariant. Sophus Lie has proved that instead to look for the
symmetry group we can first study the symmetry algebra of the
underlying PDE and then with help of  an exponential map we obtain
the corresponding symmetry group. We follow these ideas and
determine the symmetry algebra admitted by the equation.

Let us introduce now the necessary
notations.\\
 We have two sets of variables, the independent variables
 ${ t,S}$ which are in the space $X$, $(t,S)\in X$ and a dependent variable
 ${u}$ which belongs to the space $U$, i.e. $u \in U$.\\
We consider the second prolongation of the space $U$. We first
introduce the space $U_{(1)}$. It is the Euclidean space endowed
with coordinates $(u_{t}, u_{S})$ which represent all first
derivatives of $u$ with respect to $(t,S)$. Then we introduce the
space $U_{(2)}$, it is the space endowed with coordinates
$(u_{tt},u_{tS}, u_{SS})$ which represent all second derivatives
of $u$ with respect to $(t,S)$. Now we can define the second
prolongation of $U$ and a jet bundle of the order two.
\begin{definition} The $2-$nd
prolongation of $U$ is $U^{(2)}=U \times U_{(1)} \times U_{(2)}$ .
\end{definition}
\begin{definition} The $2-$nd {\bf jet bundle} of $M \subset X \times U$, or the
jet bundle of order two, is $M^{(2)}=M \times U_{(1)} \times
U_{(2)} = X \times U^{(2)}$.  $M^{(2)}$ also called $2-$nd
prolongation of $M$, it is also denoted by $pr^{(2)}M$.
\end{definition}
 We write the PDE (\ref{MainEqL3}) in the form
\begin{equation}
\Delta(w)=\Delta(x,u^{(2)})=\Delta(t,S,u,u_{t},u_{S},\dots,
u_{SS})=0,~~w \in M^{(2)},~~ u^{(2)}\in U^{(2)} \label{deltaeq}.
\end{equation}
Here $\Delta$ is a smooth map from the jet bundle $M^{(2)}$ to
some Euclidean space $\mathbb R$, i.e. $ \Delta: M^{(2)}\to
{\mathbb R} $. It defines the solution subvariety
$${ L}_{\Delta} = \{ (x,u^{(2)}): \Delta(x,u^{(2)})=0 \} \subset M^{(2)}. $$
The symmetry group $G_r$ of $\Delta$ will be defined by
\begin{equation} \label{symgroup}
G_r=\{ g \in {\rm{Diff}( M^{(2)})}{|}~~ g: ~~L_{\Delta} \to
L_{\Delta}\},~~ r=\dim G_r.
\end{equation}
The symmetry algebra $L_r={\mathcal Diff}_{\Delta}(M)$ of the PDE
$\Delta(w)=0$  can be found as a solution of the determining
equations
\begin{equation} \label{deteqn}
pr^{(2)} V(\Delta)=0 ~(mod(\Delta(w) =0)),
\end{equation}
where $pr^{(2)} V$ denotes the second prolongation of an
infinitesimal generator $V$ and has the following form
\begin{eqnarray}
pr^{(2)} V &=& \xi (S,t,u) \frac{\partial}{\partial S} + \tau
(S,t,u)
\frac{\partial}{\partial t} + \varphi(S,t,u) \frac{\partial}{\partial u}\nonumber \\
&+&\varphi^{S}(S,t,u) \frac{\partial}{\partial u_S}+\varphi^{t}(S,t,u) \frac{\partial}{\partial u_{t}}\label{ProlOper} \\
&+&\varphi^{SS}(S,t,u) \frac{\partial}{\partial
u_{SS}}+\varphi^{St}(S,t,u) \frac{\partial}{\partial u_{St}}
+\varphi^{tt}(S,t,u) \frac{\partial}{\partial u_{tt}} .\nonumber
\end{eqnarray}
Here the last five coefficients can be uniquely defined using the
first three (see \cite{olver}). The determining equations
(\ref{deteqn}) is a system of PDEs on the functions $\xi (S,t,u)$,
$\tau (S,t,u)$ and $ \varphi(S,t,u)$. The solution of this system
provides us all infinitesimal generators admitted by the studied
equation. The set of these infinitesimal generators forms a Lie
algebra which is called the Lie algebra admitted by the PDE
$\Delta(w)=0$ or the symmetry algebra of this equation. The
determining equations and the corresponding solutions to equation
(\ref{MainEqL3}) are presented in detail in \cite{Mikaelyan}.

We formulate the main result concerning the equation
(\ref{MainEqL3})
 where we do not put any constrains on the form of the function $g(\alpha)$. We
just assume that this function is a differentiable one. In this
case we obtain the following theorem.
\begin{theorem} \label{MainThm}
The equation (\ref{MainEqL3}), where $g(\alpha)$ is a
differentiable function, admits a three dimensional Lie algebra
$L_3$ spanned by the following generators
\begin{eqnarray}\label{ThreeDim}
L_3=<V_1,V_2,V_3>,~~V_1=S\frac{\partial}{\partial
S}+u\frac{\partial}{\partial u},~~ V_2=\frac{\partial}{\partial
u},~~ V_3=\frac{\partial}{\partial t}.
    \end{eqnarray}
The Lie algebra $L_3$ possesses a nonzero commutator relation
$[V_1,V_2]=-V_2.$ The Lie algebra $L_3$ has a two-dimensional
subalgebra $L_2=<V_1,V_2>$ spanned by the generators $V_1,V_2$.
The algebra $L_3$ is a decomposable Lie algebra and can be
represented as a semi-direct sum
 $L_3=L_2 \bigoplus V_3$.
\end{theorem}

The determining equations (\ref{deteqn}) admit a richer set of
solutions if the function $g(\alpha)$ has a special form. We
proved (see for details \cite{Mikaelyan}) that equation
(\ref{MainEqL3}) admits a four dimensional Lie algebra if and only
if the function $g(\alpha)$ has one of the following forms:
\begin{eqnarray} g(\alpha)&=&c_2 e^{c_1\alpha},\label{firstg}\\
g(\alpha)&=&c_2\alpha^{c_1},\label{secondg}\\
g(\alpha)&=&c_2(\rho+k \alpha)^{-\frac{1}{c_1}},~~ c_1,c_2, k -
{\rm const. } .\label{thirdg}
\end{eqnarray}
We discussed in the previous section relations between the given
form of the function $g(\alpha)$ and the corresponding utility
function. We provided also the PDEs related to the special choice
of the function $g(\alpha)$. As we mentioned before the first and
the last case, i.e. the equations (\ref{frey}) and (\ref{sipa})
were studied with the Lie group analysis before. The second case,
i.e. equation (\ref {haupt}) will be studied in
section~\ref{specsl}.

The Lie algebra $L_3$ is the symmetry algebra of equation
(\ref{MainEqL3}). Using the usual exponential map we can find the
symmetry group $G_3$ of this equation as well. It is not necessary
to find the explicit form of the symmetry group $G_3$ if we will
use invariants of this group or its subgroups to find reductions
and invariant solutions of the studied equation. It is enough to
know and to use the properties of the symmetry algebra which
corresponds to the symmetry group. The most interesting reductions
we obtain if we use one-dimensional subalgebras of $L_3$ (and
corresponding subgroups of $G_3$). The problem how to choose the
optimal system of subalgebras (and correspondingly subgroups)
which give us non-conjugate invariant solutions was solved 1982 by
Ovsiannikov (see \cite{ovsiannikov}). We follow the algorithm of
determining of an optimal system of subalgebras proposed there.

We use the notation $h_i^g$ for the subalgebras of the Lie algebra
$L_3$ and $H_i^g$ for the corresponding subgroups of the group
$G_3$. All three- and four-dimensional solvable algebras were
classified in \cite{patera&wintern}. In addition in this paper
there are provided optimal systems of corresponding subalgebras.
We use these results in this and in the next section.
\begin{proposition}\cite{patera&wintern}. ~
The optimal system of subalgebras of $L_3$ contains the
subalgebras shown in Table~\ref{TableOptSystL3}
\begin{table}[h]
\begin{center}
\begin{tabular}{|l|l|}
         \hline
         Dimension of & $~~~~~~~~~~~~~~~~~~~~~$Subalgebras\\
         the subalgebra & \\
         \hline

         1                           & $h_1^g=\left\{V_2\right\},~h_2^g=\left\{V_1\cos(\varphi)+V_3\sin(\varphi)\right\},~h_3^g=\left\{V_2+\epsilon ~V_3\right\}$\\

         2                           & $h_4^g=\big(V_2,V_3\big),~h_5^g=\big(V_1,V_3\big),~h_6^g=\big(V_1+x~ V_3,V_2\big)$\\

         \hline
\end{tabular}
\caption{ The optimal system of one- and two-dimensional
subalgebras of $L_3$, where parameters are $x\in \mathbb
R,~0\leq\phi\leq\pi,~\epsilon=\pm 1$.}\label{TableOptSystL3}
\end{center}
\end{table}
\end{proposition}

\subsection{The symmetry reductions of the general model (\ref{MainEqL3})}

There are three non-similar subalgebras of the dimension one which
we  list in Table~\ref{TableOptSystL3}. To each of these
subalgebras corresponds a one-dimensional subgroup and  they form
the set of non-conjugate subgroups. We use the optimal system of
subalgebras for invariant reductions of the equation
(\ref{MainEqL3}) to ODEs. To obtain reduced form of the PDE we
take invariant expressions as a new dependent and an independent
variables and in the new variables we obtain ODEs.
\begin{table}[ht]
\begin{center}
\begin{tabular}{|l|l|l|}
         \hline
         Subalgebra                                  & Invariants & Transformations/Reductions\\
         \hline

                                                        & &\\
         $h_1^g$
                                                               & $z=S,~W(z)=t$ & $u\rightarrow u+\rm{const}.$\\
                                                                 & &\\
                                                                  & &\\
         $h_2^g$                                         & $z=S,~W(z)=u-\epsilon t$ & $ 2 \epsilon+\frac{\sigma^2 z^2 Y'}{\left(1-z(\ln g(\rho Y))'\right)^2}=0,~Y=W',$\\
          & &\\
                                                                 &  & $g(\rho Y(z))\neq cz.$\\
                                                                  & &\\
         $h_3^g$                     & $z=\ln S-\gamma~ t~$,$~\gamma={\rm cotan}(\phi)~$ & $ 2 \gamma Y -\frac{\sigma^2(Y'+Y)}{(1-(\ln g(Y'+Y))')^2}=0,$\\
                                     &$W(z)=u S^{-1}$ &\\
                                                                 & & $~Y=W',~~g(Y+Y')\neq ce^z$.\\
                                                                  & &\\
         \hline
\end{tabular}
\caption{In the first row we list one-dimensional subalgebras from
the optimal system of $L_3$ (\ref{ThreeDim}). In the second row we
list  the corresponding invariants. In the last row we provide the
transformations or reductions of the PDE (\ref{MainEqL3}) to ODEs.
Here $\epsilon=\pm
1,~0\leq\phi\leq\pi,~\sigma^2,\rho\in\mathbb{R^+},c\in\mathbb{R}$
are parameters,  $Y',W'$ denote the differentiation of the
corresponding function on the invariant variable
$z$.}\label{TableOptSystL3}
\end{center}
\end{table}
In the first line of Table~\ref{TableOptSystL3} we see that the
first subgroup $H_1^g$ describes translations of the dependent
variable $u$ and can not be used for any reduction.

In the case of the subgroup $H_2^g$ the reduction of the given PDE
to an ODE is possible. But the solutions to the general model
(\ref{MainEqL3}) of this type are not very interesting because of
they have a trivial dependence on time (see the second row, second
line in Table~\ref{TableOptSystL3}). The most interesting case we
obtain if we use the subgroup $H_3^g$. We obtain solutions to
(\ref{MainEqL3}) of the form $u(S,t)= S W(\ln S-\gamma t)$ if we
are able to solve the first order ODE which is represented in the
last row and last line of Table~\ref{TableOptSystL3}. It is rather
impossible to present a general solution to this equation for an
arbitrary function $g(\alpha)$, but for some special choices of
$g(\alpha)$ it seems to be feasible.

\section{Symmetry properties of the special model (\ref{haupt})} \label{specsl}

We mentioned before that for some special types of the function
$g(\alpha)$ (\ref{firstg})-(\ref{thirdg}) the equation
(\ref{MainEqL3}) admits four dimensional Lie algebras.  Because
two of these cases (\ref{firstg})-(\ref{thirdg}) were studied
before we investigate now the last one, i.e. the symmetry
properties of the special model (\ref{haupt}) with the function
$g(\alpha)$ in the form (\ref{secondg}).

This equation (\ref{haupt}) contains an arbitrary constant $c_1$
but it does not include any more the parameter $\rho$. The
solution of the determining equation (\ref{deteqn}) defines four
generators of the Lie algebra $L_4$. We formulate the results in
the following Theorem.

\begin{theorem} \label{SpecThm}
Equation (\ref{haupt}) admits a four dimensional Lie algebra $$L_4
=<V_1,V_2,V_3,V_4>$$ spanned by generators
    \begin{eqnarray}\label{FourDim2}
    V_1=S\frac{\partial}{\partial S},~~
    V_2=u\frac{\partial}{\partial u},~~
    V_3=\frac{\partial}{\partial u},~~
    V_4=\frac{\partial}{\partial t}.
    \end{eqnarray}
    The Lie algebra $L_4$ possesses the following non-zero commutator relation, $[V_1,V_3]=-V_3.$
    The Lie algebra $L_4$ has a
two-dimensional subalgebra $L_2^4=<V_1,V_3>$ spanned by the
generators $V_1,V_3$. The algebra $L_4$ is a decomposable Lie
algebra and can be represented as a semi-direct sum
 $L_3=L_2 \bigoplus V_2 \bigoplus V_4$.
\end{theorem}

To provide non-equivalent reductions of the PDE (\ref{haupt})
using the Lie algebra $L_4$ we need an optimal system of one-,
two- and three-dimensional subalgebras of $L_4$. Usually just
one-dimensional subalgebras give us some interesting non-trivial
reductions. Also in the case $L_4$ we can use the classification
provided in \cite{patera&wintern}. In the paper
\cite{patera&wintern} the Lie algebra $L_4$ (\ref{SpecThm}) is
denoted by $L_4^2$.

\begin{proposition} \cite{patera&wintern}.~ \label{PropL42}
The optimal system of the one-dimensional subalgebras contains
four subalgebras, the optimal system of the two-dimensional
subalgebras contains five subalgebras and the optimal system of
three-dimensional subalgebras includes three subalgebras. All
subalgebras from these systems are listed in
Table~\ref{TableOptSystL42}.
\end{proposition}
\begin{table}[ht]
\begin{center}
\begin{tabular}{|l|l|}
         \hline
         Dimension of & $~~~~~~~~~~~~~~~~~~~~~$Subalgebras\\
         the subalgebra & \\
         \hline
         1                           & $h_1=\left\{U_3\right\},~h_2=\left\{U_1\cos\phi+U_4\sin(\phi)\right\},$\\
         & $h_3=\left\{U_2+x(U_1\cos(\phi)+U_4\sin(\phi))\right\}$,\\
         & $h_4=\left\{U_3+\epsilon(U_1\cos(\phi)+U_4\sin(\phi))\right\}$\\
         \hline
         2                           & $h_5=\left\{U_2+x(U_1\cos(\phi)+U_4\sin(\phi)),U_3\right\},$\\
                                                                 & $h_6=\left\{U_2+x(U_1\cos(\phi)+U_4\sin(\phi)),U_1\sin(\phi)-U_4\cos(\phi)\right\},$\\
                                     & $h_7=\left\{U_1,U_4\right\},$\\
                                     &$h_8=\left\{U_3+\epsilon(U_1\cos(\phi)+U_4\sin(\phi)), U_1\sin(\phi)-U_4\cos(\phi)\right\},$\\
                                     & $h_9=\left\{U_3,U_1\sin(\phi)-U_4\cos(\phi))\right\}$\\
         \hline
         3                           & $h_{10}=\big(U_2,U_1,U_4\big),~h_{11}^2=\big(U_3,U_1,U_4\big),$\\
                                                                 & $h_{12}=\big(U_2+x(U_1\cos(\phi)+U_4\sin(\phi)), U_1\sin(\phi) - U_4\cos(\phi), U_3\big)$\\
         \hline

\end{tabular}
\caption{The optimal system of the one-, two- and
three-dimensional subalgebras of $L_4$ (\ref{FourDim2}). Here the
parameters are $x\in \mathbb R,~0\leq \phi \leq \pi,~\epsilon=\pm
1$.}\label{TableOptSystL42}
\end{center}
\end{table}

\subsection{Invariant reductions of the special model (\ref{haupt})} \label{symredsection}

\textbf{Case $H_1$.} The subgroup $H_1$ related to the
one-dimensional subalgebra
    \begin{equation}\label{L42Case1}
    h_1=\left\langle\frac{\partial}{\partial u}\right\rangle
    \end{equation}
does not give rise to any reduction of the PDE (\ref{haupt})
describing the special model of the self-financing hedging
strategies in illiquid markets. It can be used to modify the
existing solutions, i.e. if we add an arbitrary constant to a
solution it will also be a solution of the same equation.

\textbf{Case $H_2$.} The subalgebra $h_2$ generates after the
exponential map a subgroup $H_2 \in G_4$. According to
Table~\ref{TableOptSystL42}  it is spanned by the generator
              \begin{equation}\label{L42Case2}
              h_2=\left\langle S\cos(\phi)\frac{\partial}{\partial S}+\sin(\phi)\frac{\partial}{\partial t}\right\rangle,~~0 \leq \phi
              \leq\pi.
              \end{equation}
             Under the action of the subgroup $H_2$
             expressions
\begin{equation}\label{invh2}
z=\ln S - \gamma~t ,~W(z)=u, ~~\gamma={\rm cotan}(\phi)\in \mathbb
R,
\end{equation}
are invariant.

{\bf Remark.} For the description of the optimal system it is
convenient to use parameters in the form $\cos(\phi)$ and
$\sin(\phi)$ to take into account that both coefficients are in
this way connected to each other. By studying invariant reductions
we can replace these parameters by $\gamma ={\rm cotan}(\phi)\in
\mathbb R$ as we did in (\ref{invh2}). The both cases
$\cos(\phi)=0,~\sin(\phi)=0$ lead to trivial invariants and
reductions and we exclude them here and further.

We use the invariant expressions (\ref{invh2}) as the new
invariant variables. Then we obtain from (\ref{haupt}) the second
order nonlinear ODE of the form
            \begin{equation}\label{L42ODE1}
             W'
            - \kappa \frac{(W''-W')W'^2}{( W''-\beta W')^2}=0,~~~\kappa=\frac{\sigma^2}
            {2 \gamma c_1^2},~\beta= \frac{c_1+1}{c_1},~c_1,~ \gamma \in\mathbb{R}\setminus \{0\}.
            \end{equation}
We excluded both cases $c_1, \gamma=0$ because these values for
the parameters lead to trivial invariants and
            equations.
The second term in equation (\ref{L42ODE1}) has a denominator. We
assume that it does not vanish. It means we exclude from the
further investigations all functions of the type
            (in the original variables $u,S,t$)
\begin{equation}\label{nulldetH2}
u(S,t)=d_1 S^{\beta}e^{-\gamma \beta t}+ d_2,~
            \beta= \frac{c_1+1}{c_1},~\gamma={\rm cotan}(\phi),~d_1,d_2\in\mathbb{R}.
            \end{equation}
Now we can multiply both terms of equation (\ref{L42ODE1}) by the
denominator and after the substitution $W'=Y$  (\ref{L42ODE1})
reduces to the ODE
\begin{equation}\label{L42Case2ODE}
(Y')^2 -Y' Y (2 \beta +\kappa) +Y^2 (\beta^2 +\kappa)
\end{equation}
This is an ODE of the first order and its solution is
\begin{eqnarray}\label{solyzH2}
Y(z)=d_1e^{k_{1,2}z},~~~~k_{1,2}=\beta +\frac{\kappa}{2}\pm
\frac{1}{2} \sqrt{\kappa(\kappa +4(\beta -1))}.
 \end{eqnarray}
The related to (\ref{solyzH2}) solutions of the original equation
(\ref{haupt}) take the form
\begin{equation}\label{SolL42Case2}
            u(S,t)=d_1~S^{k_{1,2}}e^{-\gamma k_{1,2}t }+d_2,~~~~d_1,d_2\in\mathbb{R}.
\end{equation}
\begin{figure}
\noindent
\begin{minipage}[b]{.42\linewidth}
            \centering{\includegraphics[scale=0.6]{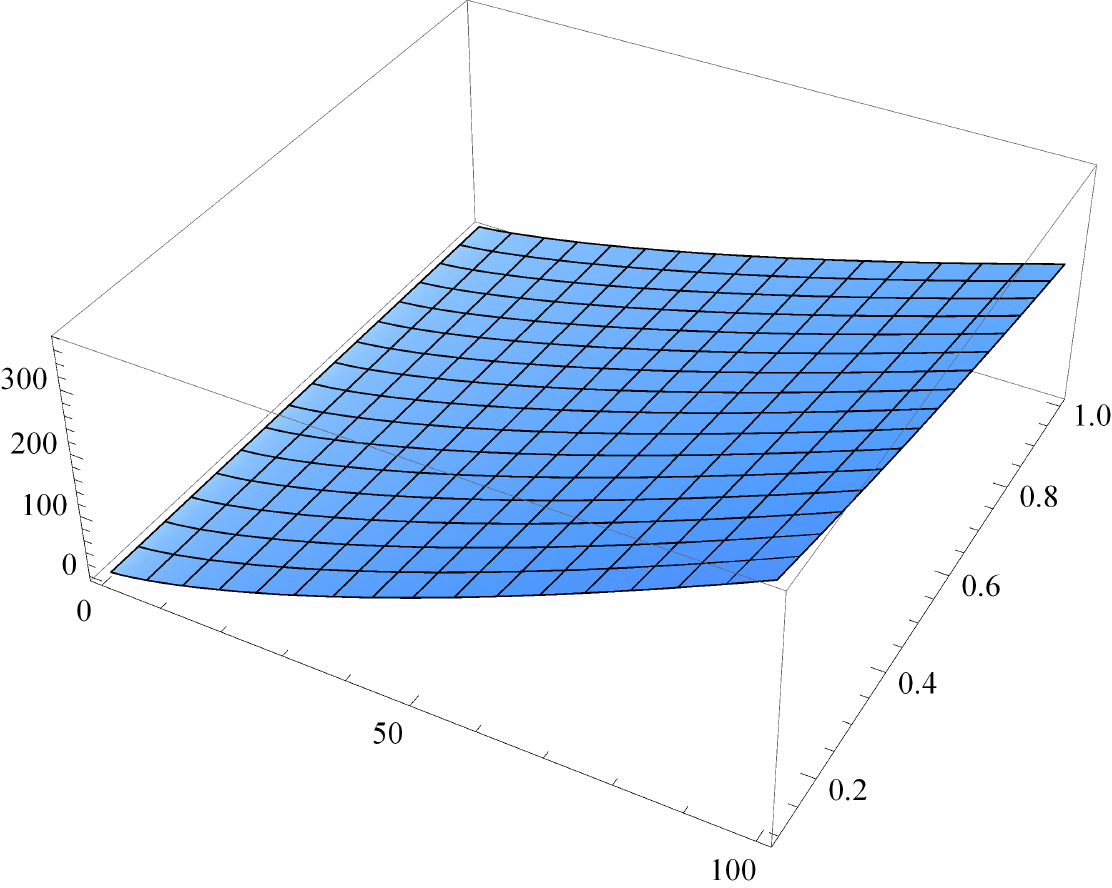}}
            \end{minipage}\hfill
\begin{minipage}[b]{.42\linewidth}
            \centering{\includegraphics[scale=0.6]{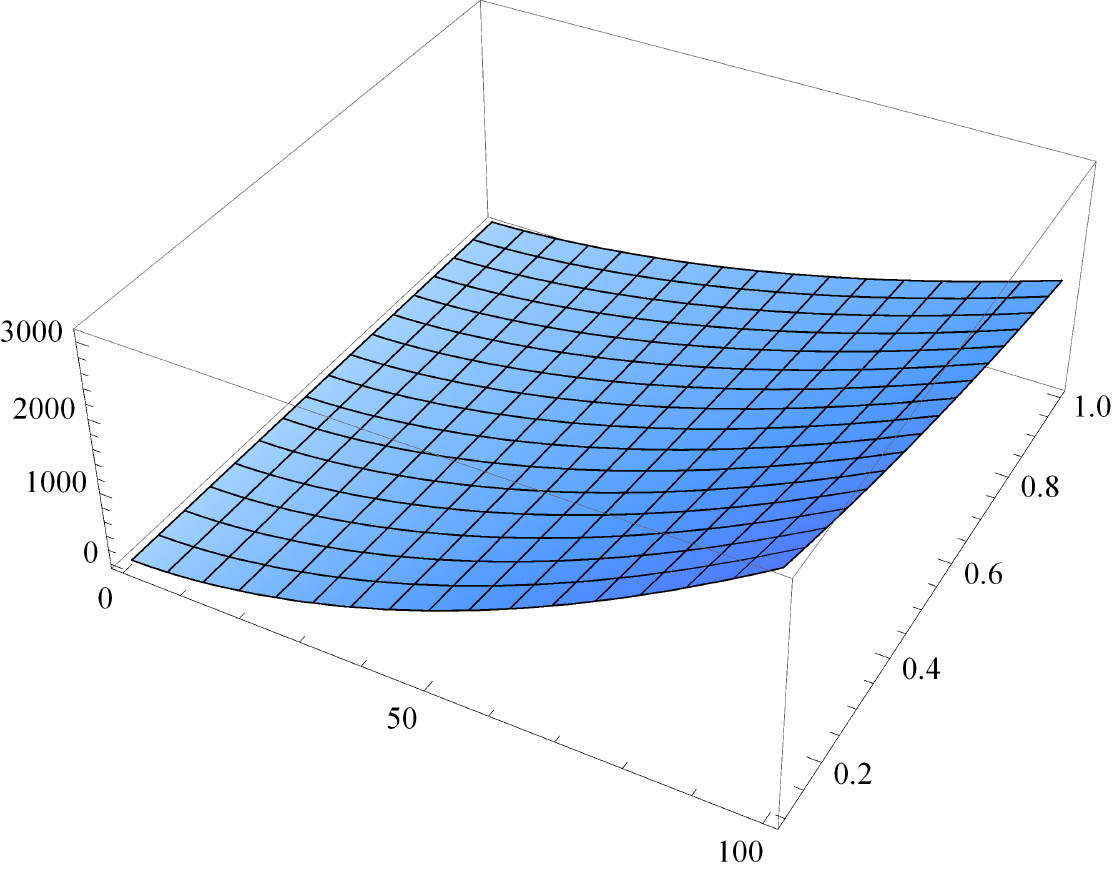}}
\end{minipage}
            \caption{Plot of the explicit solutions $u(S,t)$
            (\ref{SolL42Case2}) with $k_1=1.298$ (left panel), and $k_2=1.762$ (right
            panel) for the parameters $\beta=1.476$, $\kappa=0.11$, $\gamma=0.423$,
            $\phi=1.17,~c_1=2.1,~d_1=1,~d_2=0$. The variables $S,t$
            are
             in intervals $S \in (0.1,100)$
             and $t\in(0.1,1)$.}\label{Fig3D1}
                \end{figure}
We see that the expressions for solutions (\ref{SolL42Case2}) and
for the the functions (\ref{nulldetH2}) on which the denominator
in (\ref{haupt}) does vanish do not coincide identically.

 The solutions (\ref{SolL42Case2}) present so called {\it power}
options or futures. The payoff of these options  looks similar to
usual options, but, for example, instead of the usual payoff for a
Call option we have at expiry day $t=T$ the following payoff
$Call_{Power}(S,T)=\max[0,S^k-B]$, where $B$ is the exercise price
and $k$ is the power value.

\textbf{Case $H_3$.} The third one--dimensional subalgebra in
Table~\ref {TableOptSystL42} is spanned by the generator
                \begin{equation}\label{L42Case3}
                h_3=\left\langle u\frac{\partial}{\partial u}+x\left(S\cos(\phi)\frac{\partial}
                {\partial S}+\sin(\phi)\frac{\partial}{\partial t}\right)\right\rangle,
                ~0\leq \phi \leq\pi,x\in\mathbb{R}\setminus \left\{0\right\}.
                \end{equation}
          If the coefficient $x=0$, then the operator $U$ does not induce any reductions of
          the equation (\ref{L42Case1}).
                As the set of invariants of the corresponding group $H_3$ we take the following expressions
                $$z=\ln S- \gamma t,~W(z)=\ln u- \delta t, ~ \delta=(x\sin(\phi))^{-1},
                \gamma= {\rm cotan (\phi)}, ~ \gamma,\delta \in \mathbb{R}\setminus \{0\}.$$
                We reduce the PDE equation (\ref{haupt}) to a second order ODE using these invariants as new variables
            \begin{eqnarray}\label{L42ODE2}
            &~&\delta-\gamma W'+\frac{\sigma^2}{2 c_1^2} \frac{W'^2(W''+ W'^2-W')}{(W''+W'^2- \beta W' )^2}=0,\\
            &~&c_1, \gamma,\delta \in \mathbb{R}\setminus \{0\}\nonumber.
            \end{eqnarray}
            We multiply (\ref{L42ODE2}) by the denominator of the second term and exclude from
            the further study all the functions $W(z)$ on which the denominator vanish,
            i.e. the functions  (we present this set of functions in the original coordinates $u,S,t$)
            $$u(S,t)=d_1 S^{\beta}e^{t (\delta- \beta \gamma)} +d_2 e^{\delta t},
            ~d_1,d_2\in\mathbb{R}.$$

            After the substitution $W'=Y$, we reduce  equation (\ref{L42ODE2}) to
            \begin{eqnarray}\label{RedFin3}
(Y')^2( \delta - \gamma Y)+ Y' Y \left( 2( \delta - \gamma
Y)(Y-\beta)+\kappa Y \right) \nonumber \\
+ Y^2 \left( ( \delta - \gamma Y)(Y-\beta)^2 + \kappa Y(Y-1) \right)=0,\\
\kappa =\frac{\sigma^2}{2 c_1^2}, ~~ c_1, \gamma,\delta \in
\mathbb{R}\setminus \{0\} \nonumber.
            \end{eqnarray}

 Solutions of this first order ODE  we obtain after the integration
 \begin{eqnarray}
- 2 \int{ \frac{ ( \delta - \gamma Y) {\rm d} Y } {
 Y \left(
2( \delta - \gamma Y)(Y-\beta)+ \kappa Y
 \pm
\sqrt{ \theta Y (Y - \zeta) } \right) } }
= z+d_1, \label{integral1}\\
 \theta= \kappa( 4(\beta -1) \gamma +\kappa),
~~\zeta=\frac{4   (\beta -1) \delta}{ 4 \gamma (\beta -1)\gamma +
 \kappa},~~ \kappa =\frac{\sigma^2}{2 c_1^2}. \nonumber
 \end{eqnarray}
The integrand in (\ref{integral1}) can be transformed into a
rational function of a new variable $\tau$ by using the third
Euler substitution
\begin{equation}\label{euler}
Y= \theta \zeta (\theta - \tau^2)^{-1}.
\end{equation}
The solution of (\ref{integral1}) contains expressions for roots
of a polynomial of a degree four and is very space consuming. We
do not display this expression here. To obtain the expression for
$W(z)$ and correspondingly for $u(S,t)$, we need one additional
integration because of the substitution $Y=W'$.

\textbf{Case $H_4$.} Let us take the fourth one-dimensional
subalgebra listed in Table~\ref{TableOptSystL42}. It is spanned by
the operator
\begin{equation}\label{L42Case4}
h_4=\left\langle\frac{\partial}{\partial
u}+\epsilon\left(S\frac{\partial}{\partial
S}\cos(\phi)+\frac{\partial}{\partial
t}\sin(\phi)\right)\right\rangle,~0\leq \phi \leq\pi,\epsilon=\pm
1.
\end{equation}
We find the invariants of the corresponding Lie subgroup $H_4$ if
we solve the Lie equations, we skip this simple calculations. The
invariants in this case $H_4$ can be chosen in the form
$$z=\ln S-\gamma t,~~W(z)=u- \eta t. $$
The parameters are
\begin{equation} \label{paramH4}
\gamma=\rm cotan(\phi), ~~\eta=\frac{\epsilon }{\sin(\phi)},~~ 0<
\phi <\pi,\epsilon=\pm 1.
\end{equation}
Using the invariants (\ref{paramH4}) as new invariant variables we
obtain the following ODE of the second order from (\ref{haupt})
\begin{equation}\label{L42ODE3}
\eta- \gamma W'+\kappa \frac{(W''-W')W'^2}{(W''-\beta W')^2}=0,
~\kappa=\frac{\sigma^2}{2 c_1^2}.
\end{equation}
We exclude from the set of solutions all functions for which  the
denominator of the second term in (\ref{L42Case2ODE}) vanishes,
i.e. the family of functions (in old variables $u,S,t$)
\begin{equation}\label{detnulH4}
u(S,t)=d_1 S^{\beta}e^{-\beta \gamma t}+\eta t d_2,~~~d_1,d_2\in
\mathbb{R}.
\end{equation}
Then we multiply both terms of equation (\ref{L42Case2ODE}) with
the denominator of the second term. After the substitution $W'=Y$
we obtain the first order ODE
\begin{equation}\label{L42Case4ODE}
(Y')^2 (\eta-\gamma Y) +Y' Y (Y(2 \beta \gamma +\kappa)-2 \beta
\eta) - Y^2(Y(\beta^2 \gamma +\kappa)-\beta^2 \eta)=0.
\end{equation}
The set of solutions to equation (\ref{L42Case4ODE}) coincides
with the set of solutions to
\begin{eqnarray} \label{integralH4}
2 \int{ \frac {(\eta-\gamma Y){\rm d} Y } {\sqrt{\kappa Y ( (4
(\beta-1)\gamma +\kappa)Y - 4 (\beta-1)\eta)}} }=z+d_1.
\end{eqnarray}
To integrate the expression in (\ref{integralH4}) we use the
substitution (\ref{euler}), but now we have a different expression
for  $\zeta$. It means we use the following substitution
\begin{equation}\label{eulerH4}
Y= \theta \zeta (\theta - \tau^2)^{-1},~~ \theta=\kappa(4
(\beta-1)\gamma +\kappa),~~\zeta=  \frac{ 4 (\beta-1)\eta}{(4
(\beta-1)\gamma +\kappa)}.
\end{equation}
After the substitution (\ref{eulerH4})  equation
(\ref{integralH4}) takes the form
\begin{eqnarray}
\frac{-4 \eta}{\theta \zeta} \int{ \frac{\tau {\rm d}\tau } {b_2
\tau^2 \pm \tau +b_0}} + 4 \gamma \int{ \frac{\tau {\rm d}\tau }
{(\theta -\tau^2)(b_2 \tau^2 \pm \tau +b_0)}}
= z+d_1,\nonumber\\
b_2=2 \beta (\theta \zeta)^{-1},~~~ b_0=\kappa +2 \beta \gamma
-2\beta\zeta^{-1},~~~ d_1 \in {\mathbb R}.
\end{eqnarray}
and can be easily integrated.

\begin{figure}
\centering{\includegraphics[scale=0.8]{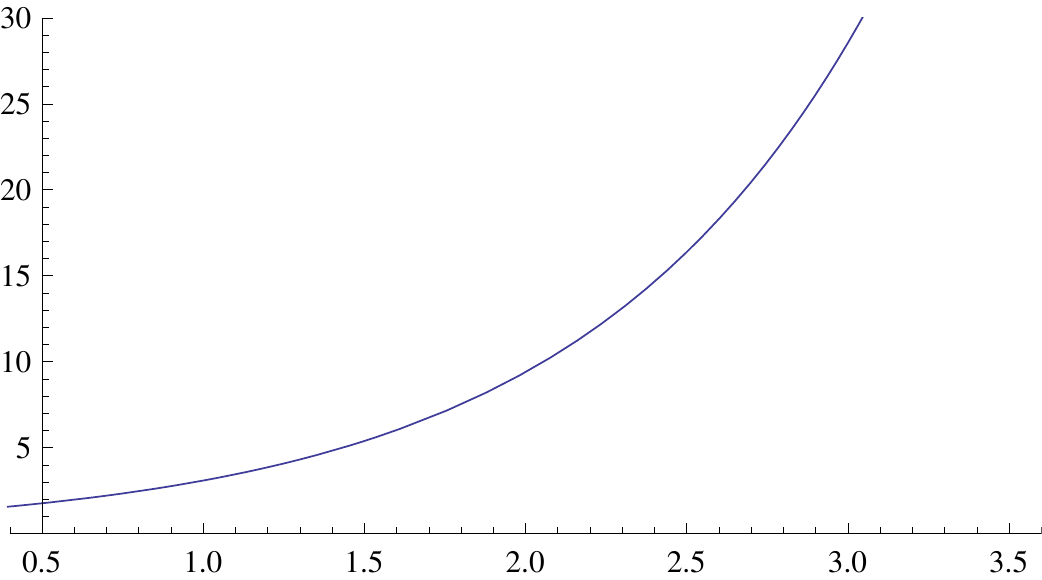}} \caption{Plot of
the solution to equation (\ref{L42Case4ODE}), where $z \in
(0.4,3.4)$ and $Y \in (0.1, 30)$. The parameters are $c_1 = 10,
\epsilon = 1, \phi = \pi/4, \sigma^2 = 0.02, ~~d_1=0
$.}\label{PicCase4Sol}
\end{figure}

We display the solution to (\ref{integralH4}) with the parameter
$\epsilon= 1$
\begin{eqnarray}
&~&2 \beta (\beta^2 \gamma+\kappa)z +d_1 =2( \beta^2 \gamma+\kappa)\ln Y \label{solplus}\\
&~&-\beta \sqrt{\theta}
\ln \left(\theta Y-\frac{1}{2} a_1  +  \sqrt{ \theta Y ( \theta Y-a_1)} \right) \nonumber\\
&~&+(\beta - 2) \kappa \ln\left(Y(\beta^2(\kappa +2
\gamma(\beta-1)))-2 \beta^2(\beta-1)\eta +\beta(\beta-2) \sqrt{ Y
( \theta Y-a_1)}\right) ,\nonumber
\end{eqnarray}
where $a_1=4 \kappa (\beta-1) \eta,~~\theta=4 \kappa 4(\beta-1)
\gamma + \kappa^2),~~d_1 \in {\mathbb R}$. The solution to
(\ref{integralH4}) with $\epsilon=-1$ has  a similar form and we
skip this expression.
 The plot of a solution to (\ref{integralH4}) for
$\epsilon =1$ is given in Figure~\ref{PicCase4Sol}.

To obtain a solution to equation (\ref{L42ODE3}) we should first
invert the function $z=f(Y)$ of the type (\ref{solplus})  and then
integrate $W(z)=\int {Y(z)}{\rm d} z$. In the general case neither
the expression (\ref{solplus}) nor the similar expression with
$\epsilon=-1$  can be inverted explicitly. But we have possibility
to solve this problem numerically and in this way solve equation
(\ref{haupt}).

\section{Conclusion}
We studied  with the Lie group analysis the general equation
(\ref{MainEqL3}) and obtain the three dimensional symmetry algebra
$L_3$ (\ref{ThreeDim}). Using the Lie algebra $L_3$ and the
corresponding optimal system of subalgebras we provide the
complete set of non-equivalent reductions of the PDE
(\ref{MainEqL3}) to ODEs. The reductions are listed in
Table~\ref{TableOptSystL3}.

In addition we were able to figure out all possible forms of the
function $g(\alpha)$ for which equation (\ref{MainEqL3}) admits an
extension of the symmetry algebra $L_3$. All such functions are
listed in (\ref{firstg})-(\ref{thirdg}). From the three equations
with the special form of the function $g(\alpha)$ two equations
were deduced earlier by different authors \cite{frey},
\cite{sircar&papanicolaou} and studied with the Lie group analysis
in \cite{bordag2010}-\cite{bordag&chmak}. The new equation in this
list is equation (\ref{haupt}) which corresponds to the function
$g(\alpha)$ (\ref{secondg}). The main properties of the symmetry
algebra $L_4$ for this equation are represented in
Theorem~\ref{SpecThm}. The invariant reductions of (\ref{haupt})
are described and studied in section~\ref{specsl}. Using, similar
to the general case, an optimal system of subalgebras we describe
the complete set of non-equivalent reductions of the PDE to
different ODEs. In all these cases we are able to solve the ODEs
and obtain solutions in exact form. We skipped in one case the
exact formula because it is too voluminous.

It is interesting to remark that the models (\ref{frey}),
(\ref{sipa}) and (\ref{haupt}) where introduced under different
finance-economical assumptions. Now we clarified how deep they are
connected from mathematical point of view. Other interesting
remark concerns the structure of the corresponding Lie algebras.
The symmetry Lie algebras of the models (\ref{frey}), (\ref{sipa})
and (\ref{haupt}) are isomorphic. It is easy to understand because
of all of them are solutions to the same determining system
(\ref{deteqn}). Less evident is why the Lie algebra admitted by
the  risk-adjusted pricing methodology model studied in
\cite{bordag20101} is also isomorphic to the previous ones. It
seems that many of the pricing options models in illiquid markets
posses similar algebraic properties.
\section*{Acknowledgements}
The authors are thankful to Magnus Larsson, the dean of the IDE
Department at Halmstad University, for the financial support of
the second author during her stay in Halmstad.

\end{document}